\documentclass[twocolumn]{aastex62}

\usepackage{graphicx}
\usepackage{dcolumn}
\usepackage{bm}
\usepackage{color}
\newcommand{\msun}{\ensuremath {\rm M}_{\odot}}
\newcommand{\yr}{\ensuremath{\rm yr}}

\newcommand{\Gpc}{\ensuremath{\,\rm Gpc}}
\newcommand{\Mpc}{\ensuremath{\,\rm Mpc}}

\newcommand{\volrate}{\ensuremath{{\rm Gpc}^{-3} {\rm yr}^{-1}}}
\newcommand{\gal}{\ensuremath{\rm gal}}

\shortauthors{Hoang, Naoz \& Kremer}

\begin{document}

\title{Neutron Star-Black Hole Mergers from Gravitational Wave Captures}

\correspondingauthor{Bao-Minh Hoang}
\email{bmhoang@astro.ucla.edu}

\author{Bao-Minh Hoang}
\affiliation{Department of Physics and Astronomy, University of California, Los Angeles, CA 90095, USA}
\affiliation{Mani L. Bhaumik Institute for Theoretical Physics, Department of Physics and Astronomy, UCLA, Los Angeles,
CA 90095}
\author{Smadar Naoz}
\affiliation{Department of Physics and Astronomy, University of California, Los Angeles, CA 90095, USA}
\affiliation{Mani L. Bhaumik Institute for Theoretical Physics, Department of Physics and Astronomy, UCLA, Los Angeles,
CA 90095}
\author{Kyle Kremer}
\affiliation{Department of Physics \& Astronomy, Northwestern University, Evanston, IL 60208, USA}
\affiliation{Center for Interdisciplinary Exploration \& Research in Astrophysics (CIERA), Northwestern University, Evanston, IL 60208, USA}
\affiliation{The Observatories of the Carnegie Institution for Science, Pasadena, CA 91101, USA}
\affiliation{TAPIR, California Institute of Technology, Pasadena, CA 91125, USA}

\begin{abstract}
LIGO's third observing run (O3) has reported several neutron star-black hole  (NSBH) merger candidates. From a theoretical point of view, NSBH mergers have received less attention in the community than either binary black holes (BBHs), or binary neutron stars (BNSs). Here we examine single-single (sin-sin) gravitational wave (GW) captures in different types of star clusters--- galactic nuclei (GN), globular clusters (GC), and young stellar clusters (YSC)--- and compare the merger rates from this channel to other proposed merger channels in the literature. There are currently large uncertainties associated with every merger channel, making a definitive conclusion about the origin of NSBH mergers impossible. However, keeping these uncertainties in mind, we find that sin-sin GW capture is unlikely to significantly contribute to the overall NSBH merger rate. In general, it appears that isolated binary evolution in the field or in clusters, and dynamically interacting binaries in triple configurations, may result in a higher merger rate.  
\end{abstract}

\section{Introduction}\label{sec:Intro}
The recent gravitational wave (GW) detections of merging BBHs and BNSs \citep{GW170817} \citep{GW150914,GW151226,GW170104,GW170608,GW170814,LIGO+18}
by LIGO/Virgo have ushered in a golden age of GW astrophysics. The first (O1) and second (O2) observing runs of the advanced LIGO/Virgo detector network have yielded inferred estimates for the merger rate of BBHs ($9.7-101~\volrate$) and BNSs ($110-3840~\volrate$) \citep{Abbott+19}. A very diverse range of mechanisms and astrophysical environments have been invoked to explain these mergers, such as: dynamical interactions in globular clusters \citep[e.g.][]{Portegies+00,Wen,OLeary+06,Antonini+14,Rodriguez+16,Oleary+16,Kremer+19} and galactic nuclei \citep[e.g.][]{OLeary+09,Kocsis+Levin,AP12,Ramirez-Ruiz+15,Hoang+18,Fernandez+Kobayashi19}, active galactic nuclei \citep[e.g.][]{McKernan+12,Bartos+17,Stone+17,Secunda+19}, isolated binary evolution in the field \citep[e.g.][]{Mandel+deMink,deMink+Mandel,Belczynski+16,Marchant+16}, Population III stars \citep[e.g.][]{Kinugawa+14,Kinugawa+16,Hartwig+16,Inayoshi+16,Dvorkin+16}, and primordial black holes \citep[e.g.][]{Bird+16,Clesse+17,Sasaki+16}.

The non-detection  of any neutron star-black hole (NSBH) mergers during O1 and O2 puts a $90\%$ confidence interval upper limit on the NSBH merger rate of $0-610~\volrate$. However, there are currently several candidates for NSBH mergers in O3 and it is likely that there will be a confirmed detection within the decade (see \citet{GraceDB}). NSBHs are much less well studied than either BBHs or BNSs. However, over the years there have been a number of studies estimating the rate of NSBH mergers from a number of channels. In this work we explore a relatively unexplored channel for producing NSBH mergers --- eccentric GW captures in dense clusters. GW captures are well-studied as a promising formation mechanism for BBHs, particularly in GN. NSBH captures are less straightforward than BBH captures due to the potential presence of tidal effects. Furthermore, NSs and BHs interact with each other less often than they do with members of their own species due to mass segregation in clusters. We explore these complications in three different types of cluster in this work: GNs, GCs, and YSCs. We then compare our results to results from other channels, consider the limits of the different channels, and finally discuss likely origins for a future NSBH merger detection in LIGO. We begin by summarizing the various channels that have been proposed to explain NSBH mergers below:
\begin{enumerate}
    \item \textit{Mergers in the Field}: The most well studied channel for merging NSBH is binary evolution in the field. Early studies using population synthesis have resulted in a large range of merger rates, $0.68-42.8~\yr^{-1}$ in aLIGO, due to many uncertainties in binary evolution models \citep[e.g.,][]{Sipior+02,Pfahl+05,Belczynski+07,Belczynski+10,OShau+10}. \citet{Belczynski+11} studied X-ray source Cyg X-1--- a likely NSBH progenitor--- and found an ``empirical" rate of NSBH mergers  based on the evolution of this system: $(2 - 14)\times 10^{-10} {\rm yr}^{-1} \gal^{-1}$ (inferred volumetric rate $\sim 0.002 - 0.014~\volrate$), which is much smaller than the previously estimated rates from population synthesis. A more recent binary population synthesis study by \citet{Dominik+15} estimated NSBH merger rates in aLIGO to be $0.6-1.2 \yr^{-1}$ for a standard binary evolution model, up to $3.6-5.7~\yr^{-1}$ for an optimistic common envelope evolution model, and down to $0.03-0.3~\yr^{-1}$ for a pessimistic model with high BH kicks (inferred volumetric rate across all models $\sim 0.1-14~\volrate$). More recently, \citet{Kruckow+18} found an optimistic upper limit to NSBH mergers via isolated binary evolution in the field of up to $\sim 53~\volrate$.

    \item \textit{Mergers in Globular Clusters (GC)}: 
    NSs have been well-observed in GCs dating back to the 1970s as both X-ray \citep[e.g.][]{Clark1975,Heinke+05} and radio sources \citep[e.g.][]{Lyne+87,Ransom08}. Over the past decade, a growing amount of evidence has suggested GCs also retain BH populations \citep[e.g.][]{Strader+12,Giesers+19}. Thus the question of NSBH formation in GCs arises naturally. Several studies have suggested that the rate of NSBH mergers is much lower in GCs than in the field through various lines of reasoning \citep{Phinney91,Grindlay+06,Sadowski+08}. A few authors have since calculated the rates of NSBH mergers in GC through dynamical processes. For example, \citet{Clausen+13} studied binary-single (bin-sin) interactions in a static cluster potential, and found a merger rate of $\sim 0.01 - 0.17~\volrate$. More recently, \citet{Ye+20} studied dynamically formed NSBHs in GCs using the code CMC (Cluster Monte Carlo, see \citet{Kremer+19} for details), and found a rate of $0.009 - 0.06~\volrate$. \citet{ArcaSedda20} studied hyperbolic bin-sin interactions in dense clusters, and found a NSBH merger rate of $3.2 \times 10^{-3} - 0.25~\volrate$ in GCs. The upper limits of these GC rates are indeed only comparable to the lower limits of the theoretical field rates, supporting the idea that field channels dominate over GC channels for NSBH mergers. Note that all three of the aforementioned studies focused on \textit{binary-mediated} interactions, i.e. binary-single (bin-sin), and binary-binary (bin-bin) interactions, and did not include \textit{single-single (sin-sin)} encounters, which will be the focus of this paper.
    
    \item \textit{Mergers in Galactic Nuclei (GN)}: The extent of previous studies for NSBH mergers in GN are similarly limited. 
    For example, \citet{ArcaSedda20} studied bin-sin mergers in GN and found a rate of $\sim 9 \times 10^{-3} - 1.5 \times 10^{-2}~\volrate$ .\citet{OLeary+09} focused on mergers of BBHs in GN resulting from sin-sin GW captures \citep[e.g.,][]{Quinlan+Shapiro87,Lee93}, using compact object densities resulting from Fokker-Planck simulations. They estimated that the rate of NSBH mergers from this channel will be roughly $10^{-11} \yr^{-1} \gal^{-1}$ for a GN around a $4\times 10^6~\msun$ SMBH, or 1\% of the BBH rate. \citet{Tsang13} also estimated the rate of NSBH mergers in GN, but using density profiles of an isothermal sphere instead of density profiles from a Fokker-Planck simulation, and found a merger rate of $\sim 7 \times 10^{-11} - 9 \times 10^{-10} \yr^{-1} \gal^{-1}$ for a GN surrounding a $4\times 10^6~\msun$ SMBH (calculated from their Eq. A9).
    
    We note that there is a great deal of subtlety and uncertainty concerning the conversion of a per galaxy merger rate for fixed SMBH mass to a volumetric/expected detection rate for GW captures in GN. The most straightforward way is to simply multiply the per galaxy rate by a galaxy number density in the universe to find the volumetric rate; and then multiply the volumetric rate by the volume observable by LIGO to find an expected detection rate. However, as \citet{OLeary+09} noted, there may be significant variance in central cusp densities between different GNs with the \textit{same SMBH mass}. Since the rate of GW captures scales as density squared, this variance may lead to an enhancement of the aforementioned  volumetric/expected detection rate by a factor of $\xi$. The true value of $\xi$ is highly uncertain. Whereas \citet{OLeary+09} and \citet{Kocsis+Levin} estimate $\xi$ to be $\gtrsim 30$, \citet{Tsang13} found that $\xi$ is at most $\sim 14$ under very optimistic assumptions. \citet{OLeary+09} and \citet{Tsang13} found volumetric (expected LIGO detection) rates of $\sim 0.07~\volrate$ ($\sim 1~\yr^{-1}$) and $\sim 0.05-0.6~\volrate$ ($\sim 1.6-20~\yr^{-1}$), respectively, using their different values of $\xi$. Without enhancement from $\xi$, these rates will decrease to roughly $\sim 0.002~\volrate$ ($\sim 0.03~\yr^{-1}$) and $\sim 0.004-0.05~\volrate$ ($0.1-1.5~\yr^{-1}$), respectively. Due to the high uncertainty in the value of $\xi$, in this work we calculate and adopt GW capture rates \textit{without $\xi$} as our fiducial rates, which yields conservative estimations. However, we will discuss the implications for LIGO should $\xi$ be significant.
    
    Aside from GW captures, GN can be the site of binary mergers induced by interactions with the SMBH \citep[e.g.,][]{AP12,Naoz16,Stephan+16,Stephan+19,Hoang+18}, via the Eccentric Kozai-Lidov Mechanism \citep[EKL][]{Kozai,Lidov,Naoz16}.  Recently \citet{Lu+19} suggested that supernova natal kicks can tend to shrink the post supernova separation. Moreover they showed that these systems are more likely to stay in a triple configuration near an SMBH. On the other had, a supernova kick rarely keeps stellar-mass tertiary. Thus, torques from the SMBH can lead to enhancement of NSBH mergers, compared too field binaries.

    Subsequently, \citet{Stephan+19} studied stellar binary evolution in GN with EKL  including self-consistent post-main sequence stellar evolution and found that LIGO may detect NSBH mergers from this mechanism at a rate of $2-5\yr^{-1}$ (inferred volumetric rate: $0.17-0.33~\volrate$). \citet{Fragione+19a} performed a study of compact binary mergers in GN induced by EKL, considering different binary parameter distributions and SMBH masses, and found a NSBH merger rate of $0.06-0.1~\volrate$. Comparing these EKL rates to the sin-sin GW capture rates from \citet{OLeary+09} and \citet{Tsang13}, we see that if $\xi$ is small (i.e. the variance in GN density is low), mergers induced by EKL will dominate over mergers from GW captures in GN. Conversely, if $\xi$ is significant, then mergers from GW captures will be either comparable to or dominate over mergers from EKL. 
    
    Recently \citet{McKernan+20} studied compact object binary mergers in AGN disks, the gas in which has previously been shown to potentially accelerate binary mergers \citep[e.g.][]{McKernan+12,Bartos+17,Stone+17,Secunda+19}. They found that this channel can potentially produce NSBH mergers at rates of $f_{\rm AGN,BBH} (10-300)~\volrate$, where $f_{\rm AGN,BBH}$ is the fraction of BBH mergers observed by LIGO that come from the AGN channel. They expect that 10-20\% of NSBH mergers from this channel will have electromagnetic counterparts, which can help disentangle this channel from others.
    
     \item \textit{Mergers in Young Stellar Clusters (YSC)}: Most stars, including massive stars that are BH and NS progenitors, are born in YSCs \citep{Carpenter2000,Lada2x03,Porras+03}. Their higher density relative to the galactic field means that compact binaries can form from dynamical interactions similar to those in GCs, as well as from stellar binary evolution. As a result, a number of studies have explored YSCs as a possible birthplace for BBHs  \citep[e.g.][]{PZ+M02,Banerjee+10,Kouwenhoven+10,Goswami+14,Ziosi+14,Mapelli16,DiCarlo+19,Banerjee17,Banerjee18,Fujii+17,Kumamoto+19,Rastello+19}, with promising results. Recently, \citet{Rastello+20} studied the formation of NSBHs in YSCs from redshifts 0-15. They found that YSCs can produce NSBHs that merge in the local universe (redshift $< 0.1$) at a rate of $\sim 28~\volrate$, through a combination of binary evolution and dynamical interactions. Most of these NSBHs will be ejected from YSCs before they merge, and so will ostensibly be ``field" binaries when they are detected by LIGO. However, NSBHs that formed in YSCs have a different mass spectrum from those that formed in true isolation in the field, and may be differentiated in this way. The rate found in \citet{Rastello+20} is likely an optimistic estimation of the YSC merger rate, due to the following reasons. \citet{Rastello+20} assume a NS natal kick distribution with a root mean square of $15~{\rm km/s}$, whereas observational studies of pulsar proper motions in the literature show that a majority of NSs likely receive very large natal kicks ($\sim 200-500~{\rm km/s}$) at birth \citep[e.g.][]{Hansen+Phinney97,Lorimer+Bailes97,Cordes+Chernoff98,Fryer+99,Hobbs+04,Hobbs+05,Beniamini+Piran16}. High velocity natal kicks tend to disrupt binaries and may significantly reduce the rate of NSBH formation from binary evolution. In addition, the high stellar densities and fractal initial conditions adopted in \citet{Rastello+20} may not be representative of all YSCs, and therefore may overestimate the influence of dynamis. For comparison, lower density models found in another recent work, \citet{Fragione+Banerjee20}, resulted in an upper limit of $3 \times 10^{-3}~\volrate$ for the NSBH merger rate from binary evolution and dynamical exchanges in YSCs. Note that while the analysis in \citet{Rastello+20} and \citet{Fragione+Banerjee20} included dynamical binary interactions and exchanges, they also did not include sin-sin GW capture. We will give an order of magnitude upper limit estimation of the rate of NSBH mergers due to sin-sin GW captures in YSCs in this work.
    
    \item \textit{Mergers in Triples:} Stellar multiplicity studies have shown that $\sim 15\%$ of massive stars --- progenitors of BHs and NSs --- have at least two stellar companions \citep[e.g.][]{Raghavan+10,Sana+13,Dunstall+15,Jimenez+19}. Several studies have explored the formation of BBH mergers in stellar triples and quadruples \citep[e.g.][]{Antonini+17,Silsbee+Scott17,Fragione+Kocsis19,Liu+Lai19}. Recently, \citet{Fragione+Loeb19a} and \citet{Fragione+Loeb19b} studied NSBH mergers in field triples and found merger rates of $\sim 1.9 \times 10^{-4}-22~\volrate$, where the wide range comes from uncertainties in the metallicity of the progenitor population, and the magnitude of BH and NS natal kick velocities.

\end{enumerate}

The paper is organized as follows: We begin with describing the basic equations that govern sin-sin GW capture in Section \ref{sec:sin-sin}. We then calculate the sin-sin NSBH merger rate in GCs, GNs, and YSCs in Section \ref{sec:rates}. Finally, we offer our discussions about the most probably NSBH merger channels in Section \ref{sec:Dis}.

\section{Single-Single Gravitational Wave Captures}\label{sec:sin-sin}
Two compact objects undergoing a close encounter can emit enough gravitational wave energy to become a bound binary. Because these encounters are relativistic, and the velocity dispersion of galactic nuclei and other clusters are much less than the speed of light, they are almost always nearly parabolic \citep{Quinlan+Shapiro87,Lee93}. We consider these approximately parabolic encounters and subsequent gravitational wave captures of stellar-mass black holes of mass $m_{\rm BH}$ and neutron stars of mass $m_{\rm NS}$,  total mass $M_{\rm tot} = m_{\rm BH} + m_{\rm NS}$, a symmetric mass ratio $\eta = m_{\rm BH} m_{\rm NS}/((m_{\rm BH} + m_{\rm NS})^2)$, a relative velocity of $v_{\rm rel}$, and an impact parameter of $b$. The energy that is emitted in GWs in such an encounter is \citep{Peters+63,Turner77}:
\begin{equation}\label{eq:delE}
\Delta E_{\rm GW} = -\frac{85 \pi G^{7/2}}{12 \sqrt{2} c^5}\frac{\eta^2 M^{9/2}_{\rm tot}}{r^{7/2}_{\rm p}}
\end{equation}
where $c$ is the speed of light, $G$ is the gravitational constant, and $r_{\rm p}$ is the distance of closest approach of the encounter:
\begin{equation}
r_p = \Bigg(\sqrt{\frac{1}{b^2} + \frac{G^2 M^2_{\rm tot}}{b^4 v^4_{\rm rel}} + \frac{G M_{\rm tot}}{b^2 v^2_{\rm rel}}}\Bigg)^{-1}.
\end{equation}
If $|\Delta E_{\rm GW}| > \frac{1}{2}M_{\rm tot} \eta v^2_{\rm rel}$ (the kinetic energy of the encounter), a bound NSBH binary is formed \citep[e.g.,][]{Lee93}. This criterion implies a maximum impact parameter $b_{\rm max}$ to form a bound binary \citep[e.g.,][]{OLeary+09,Gondan+18a}:
\begin{equation}\label{eq:bmax}
    b_{\rm max} = \left(\frac{340 \pi \eta}{3}\right)^{1/7} \frac{G M_{\rm tot}}{c^2} \left(\frac{v_{\rm rel}}{c}\right)^{-9/7} \ .
\end{equation}
There is also a minimum impact parameter $b_{\rm min}$ to form a bound binaries, as encounters with $b < b_{\rm min}$ will result in a direct collision rather than a bound binary. These encounters may result in an electromagnetic event but will likely not result in any strong GW signals. This collisional impact parameter is defined as \citep{Gondan+18a},
\begin{equation}\label{eq:bmin}
    b_{\rm min} = \frac{4 G M_{\rm tot}}{c^2}\left(\frac{v_{\rm rel}}{c}\right)^{-1}.
\end{equation}
The total GW capture cross section is thus:
\begin{equation}
    \sigma(m_{\rm BH}, m_{\rm NS},v_{\rm rel}) = \pi(b^2_{\rm max} - b^2_{\rm min}).
\end{equation}

\begin{figure}
    \centering
    \includegraphics[width=\linewidth]{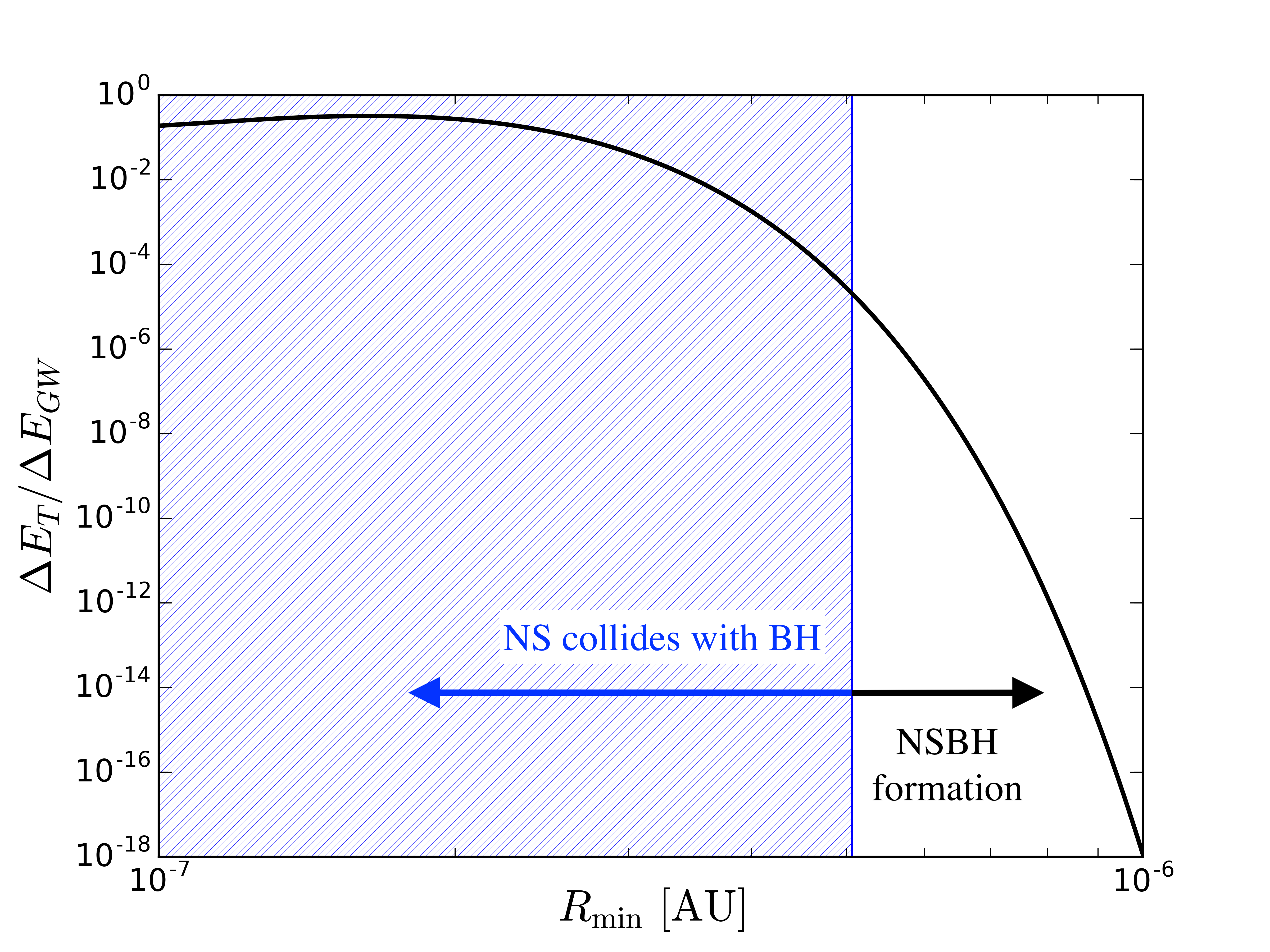}
    \caption{\textbf{Ratio of energy lost to tides to energy lost to GWs ($\Delta E_{\rm T}/\Delta E_{\rm GW}$), as a function of encounter periastron ($R_{\rm min}$)}. Plotted for parabolic encounters between a $5~\msun$ BH and a $1.4~\msun$ NS. The region highlighted blue denotes encounters with impact parameter less than $b_{\rm min}$ (given by Eq. (\ref{eq:bmin}))-- these encounters will result in a direct collision between the BH and NS. The region of interest for us is the region to the right of the blue line, where a NSBH can form. In this region, $\Delta E_{\rm T}/\Delta E_{\rm GW} < 10^{-4}$, an extremely small value. We note that encounters between a NS and a BH of mass greater than $5~\msun$ will result in even smaller values of $\Delta E_{\rm T}/\Delta E_{\rm GW}$. Thus, we conclude that we can ignore tidal effects in our calculations. }
    \label{fig:TidesvGW}
\end{figure}

We note that during these encounters, energy is also lost due to tidal oscillations in the neutron star excited by the black hole \citep[e.g.,][]{PT77}, and contributes to $\sigma$. To check whether we should include this effect in our calculations or whether it can be safely neglected, we approximate the tidal energy dissipated during a parabolic encounter according to the formalism presented in \citet{PT77}:
\begin{equation}
    \Delta E_{\rm T} =  \Big(\frac{G m^2_{\rm NS}}{R_{\rm NS}}\Big)\Big(\frac{m_{\rm BH}}{m_{\rm NS}}\Big)^2 \sum_{l = 2,3,...}\Big(\frac{R_{\rm NS}}{R_{\rm min}}\Big)^{2 l + 2} T_l,    
\end{equation}
where $R_{\rm NS}$ is the radius of the neutron star, $R_{\rm min}$ is the periastron of the approach, and $T_l$ are dimensionless values associated with each spherical harmonic $l$ (see \citet{PT77} for calculation of $T_l$). We only consider the quadrupole mode ($l = 2$), which dominates over the other modes \citep{PT77}. We approximate the NS as a polytropic star of index $n = 0.5$ \citep[e.g.,][]{Finn87}, and use values from Table 1 of \citet{KS95} to aid in the calculation of $T_l$. Note that since there is a minimum impact parameter, there is minimum possible value of $R_{\rm min}$. For a parabolic encounter the relationship between the impact parameter and the periastron distance is:
\begin{equation}\label{eq:Rmin}
    R_{\rm min}(b) = \frac{b^2 v^2_{\rm rel}}{2 G M_{\rm tot}} \ .
\end{equation}
Thus, combining Equations (\ref{eq:bmin}) and (\ref{eq:Rmin}), we find the minimum possible $R_{\rm min}$ to be:
\begin{equation}
    R_{\rm min}(b_{\rm min}) = \frac{8 G M_{\rm tot}}{c^2} \ .
\end{equation}
Encounters with $R_{\rm min} < R_{\rm min}(b_{\rm min})$ will result in a direct collision between the BH and NS instead of a bound binary. In Figure \ref{fig:TidesvGW} we plot $\Delta E_{\rm T}/\Delta E_{\rm GW}$--- the ratio of energy lost to tidal oscillations to the energy lost to gravitational waves--- as a function of $R_{\rm min}$, for a $5~\msun$ BH and a $1.4 \msun$ NS. We have also marked the region where $R_{\rm min} < R_{\rm min}(b_{\rm min})$. We see that in the region of interest where $R_{\rm min} > R_{\rm min}(b_{\rm min})$, i.e., where bound binaries can form, $\Delta E_{\rm T}/\Delta E_{\rm GW} < 10^{-4}$, an extremely small value. We have verified (not shown to avoid clutter), that for larger BH masses, $\Delta E_{\rm T}/\Delta E_{\rm GW}$ is even smaller. This is consistent with previous studies about NS-NS captures \citep[e.g.,][]{Gold+12,Chirenti+17}. Thus, we can safely neglect tides in our calculation of the capture cross section.

\section{Event Rates}\label{sec:rates}
The rate of NSBH binary formation for a single cluster is:
\begin{eqnarray}\label{eq:fullrate}
    \Gamma_{\rm cl} =& &\int {\rm d}r~4 \pi r^2 n_{\rm BH}(r) n_{\rm NS}(r)\nonumber \\ 
    \times & &\int {\rm d} m_{\rm BH}~\mathcal{F_{\rm BH}}(m_{\rm BH}) \int {\rm d} m_{\rm NS}~\mathcal{F_{\rm NS}}(m_{\rm NS}) \nonumber \\
    \times & & \int {\rm d}v_{\rm rel}~\psi_{ m_{\rm BH},m_{\rm NS}}(r,v_{\rm rel}) \sigma v_{\rm rel},
\end{eqnarray}
where $n_{\rm BH}(r)$ and $n_{\rm NS}(r)$ are the number densities of a black holes and neutron stars, respectively; $\mathcal{F_{\rm BH}}(m_{\rm BH})$ and $\mathcal{F_{\rm NS}}(m_{\rm NS})$ are the mass probability distributions for black holes and neutron stars, respectively; $\psi_{ m_{\rm BH},m_{\rm NS}}(r,v_{\rm rel})$ is the distribution of the relative velocity between $m_{\rm BH}$ and $m_{\rm NS}$ at r.

For GCs, we approximate the BH and NS populations of each cluster as following a Maxwellian velocity distribution. Thus, the BH and NS populations have velocity dispersions of $v_{\rm d,BH}$ and $v_{\rm d, NS}$, respectively. We then calculate the average relative velocity between BHs and NSs in a cluster, $<v_{\rm rel}> = \sqrt{(8/\pi)(v^2_{\rm d,BH}+v^2_{\rm d,NS})}$, and use this constant value in Equation (\ref{eq:fullrate}). Note that in this calculation we have neglected any mass or $r$ dependence in $v_{\rm rel}$ for simplicity. Thus we have:
\begin{equation}\label{eq:GCvrel}
 \int_{\rm GC} {\rm d}v_{\rm rel}~\psi_{ m_{\rm BH},m_{\rm NS}}(r,v_{\rm rel}) \sigma v_{\rm rel} \approx \sigma <v_{\rm rel}>.
\end{equation}   

For GNs, \citet{OLeary+09} showed that the last integral in Equation (\ref{eq:fullrate}) is only weakly dependent on the relative velocity distribution, and is well approximated by:
\begin{equation}\label{eq:GNvrel}
\int_{\rm GN} {\rm d}v_{\rm rel}~\psi_{ m_{\rm BH},m_{\rm NS}}(r,v_{\rm rel}) \sigma v_{\rm rel} \approx \sigma v_c(r),
\end{equation}
where $v_c(r) = \sqrt{G m_{\rm SMBH}/r}$ is the circular velocity at $r$ and $\sigma$ is evaluated at $v_{\rm rel} = v_c(r)$. 

We can then calculate the nominal volumetric NSBH merger rate due to sin-sin GW capture:
\begin{equation}
    \Gamma_{\rm NSBH} = n_{\rm cl} \Gamma_{\rm cl},
\end{equation}
where $n_\mathrm{cl}$ is the density of clusters (either GN or GC, we use a slightly different calculation for YSCs, see section \ref{YSC}) in the local universe and $\Gamma_{\rm cl}$ is the rate per cluster. Note that even though the capture rate is not technically the same as the merger rate, the vast majority of binaries that form due to GW capture tend to be very tight, eccentric, and merge very quickly after capture. Thus, the capture rate is an extremely good approximation of the merger rate \citep[e.g.][]{OLeary+09,Gondan+18a}.

\begin{figure}
    \centering
    \includegraphics[width=\linewidth]{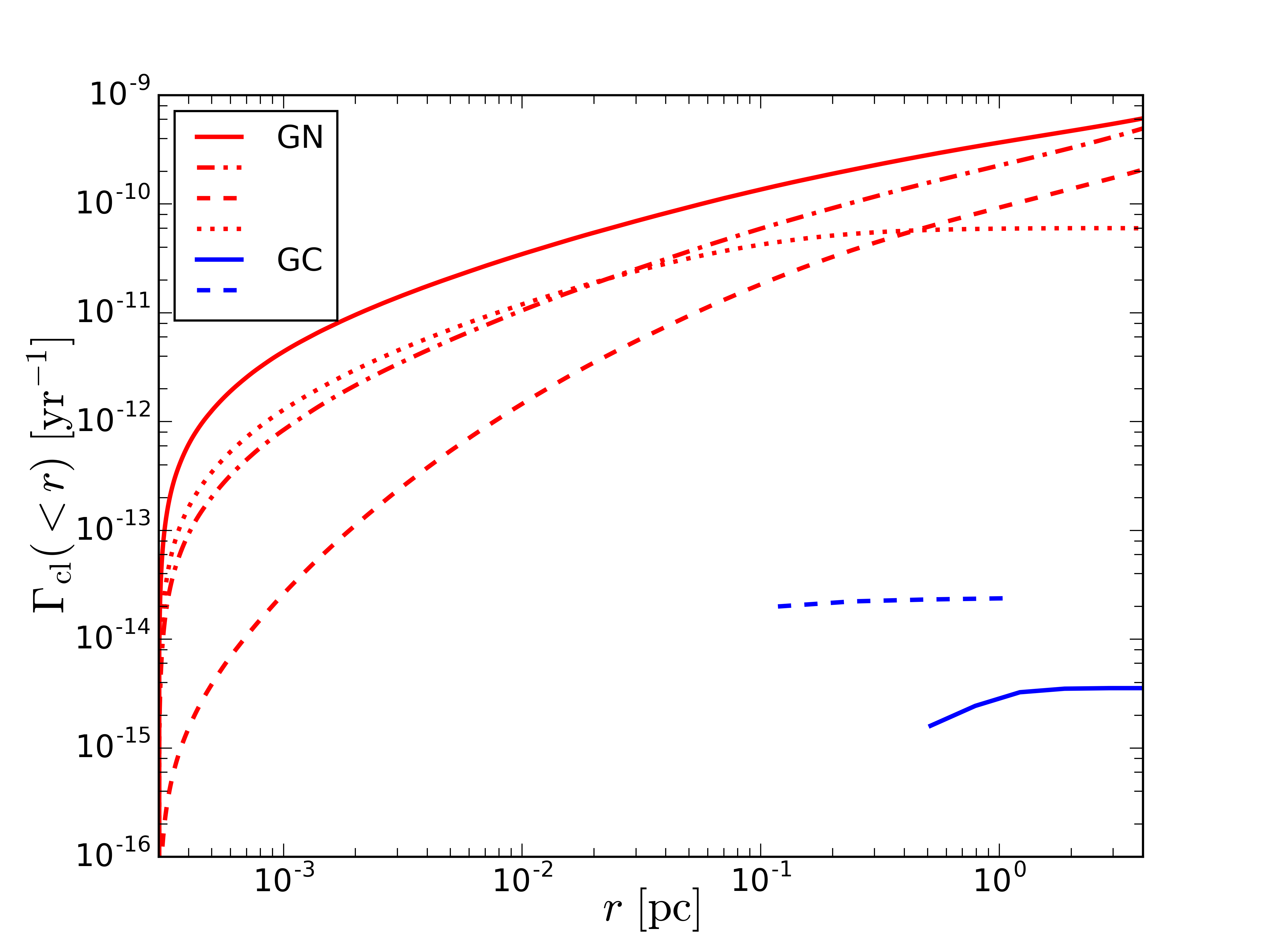}
    \caption{\textbf{Cumulative merger rate per cluster $\Gamma_{\rm cl}$ ($\yr^{-1}$) as a function of $r$}. Note that for GN cases (red), $r$ denotes the distance from the SMBH; whereas for GC cases (blue), $r$ denotes the distance away from GC center. The GC lines do not extend below $r \sim 0.1~{\rm pc}$ because our GC models do not show NSs in the inner regions of the cluster, see Section \ref{GC} for more details. For GN cases, the four different curves correspond to four different GN evolution models from Figures 1 \& 2 of AP16 (see Section \ref{GN} for a short description of Models 1-4, and AP16 for full details): solid --- Model 1, dash dot --- Model 2, dashed --- Model 3, dotted --- Model 4. For GC cases, the solid line corresponds to the 1 Gyr case, and the dashed line corresponds to the 10 Gyr case.}
    \label{fig:CumuRates}
\end{figure}

\subsection{Globular Clusters (GC)}\label{GC}
We calculate the capture rates for a single simulated cluster that is representative of a typical Milky Way cluster (initial cluster mass $4 \times 10^5~\msun$, final mass $2 \times 10^5~\msun$, core radius $\sim 1~{\rm pc}$, galactocentric distance of $8\,$kpc with Milky Way-like potential, and metallicity of $0.01Z_{\odot}$), taken from the latest CMC catalogue \citep{Kremer+19}. We compare the contribution from younger clusters versus older clusters by considering this simulated cluster at $1~{\rm Gyr}$ and $10~{\rm Gyr}$. To perform the integral in Equation (\ref{eq:fullrate}), we numerically calculate the densities $n_{\rm BH}(r)$ and $n_{\rm NS}(r)$, and the mass distributions $\mathcal{F_{\rm BH}}(m_{\rm BH})$ and $\mathcal{F_{\rm NS}}(m_{\rm NS})$ from the simulation data (note this particular simulation contains 527 BHs and 825 NSs at $t=1\,$Gyr and contains 38 BHs and 778 NSs at $t=10\,$Gyr). As previously mentioned in this section, we adopt Maxwellian velocity distribution for the BH and NS populations. We fit the BH and NS velocities with respect to cluster center to a Maxwellian distribution in order to find their velocity dispersions, and from those velocity dispersions calculate the average relative velocity $<v_{\rm rel}>$ (see Equation (\ref{eq:GCvrel})).  In Figure \ref{fig:CumuRates}, we show the cumulative merger rate of a cluster as a function of distance, $r$, from the center for the young  (blue, solid line) and old (blue, dashed line) cluster.

We find a total merger rate per GC of $\Gamma_{\rm cl, GC} \sim 4 \times 10^{-15}~\yr^{-1}$ ($2 \times 10^{-14}~\yr^{-1}$) for the $1~{\rm Gyr}$ ($10~{\rm Gyr}$) GC. In the younger $1$~Gyr cluster, the BH population contains many higher mass BHs (mass range $5-40$~$\msun$). Through mass segregation, this BH population forms a dense subsystem in the cluster's center that subsequently generates significant energy through ``BH burning'' \citep[the cumulative effect of dynamical binary formation, hardening, and ejections; for review, see][]{Kremer+20a}. This process  influences the large-scale structural properties of the host cluster, in particular, delaying the onset of cluster core collapse by preventing the migration of lower-mass stars, including NSs, to the cluster's center \citep[e.g.,][]{Merritt+04,Mackey+08,BreenHeggie13,Peuten+16,Wang+16,ArcaSedda+18,Kremer+19a,Ye+20}. As a result, there is only density overlap between BHs and NSs in the outer regions of the cluster where densities are low (i.e., there are no NSs in the inner most regions where the BHs dominate). This results in an extremely low rate of capture.

However, as the cluster ages, the total number of retained BHs decreases as a result of the dynamics within the BH subsystem \citep[see, e.g.,][]{Morscher+15}. Additionally, because the highest mass BHs are dynamically ejected first, as the cluster ages, the BH mass distribution becomes increasingly dominated by lower-mass BHs  ($5-15$~$\msun$). As a consequence of these effects, the energy generated through the BH burning process (and therefore effect of the BHs on the clusters radial profile) becomes less significant as the cluster ages. Thus, the NS population is able to infiltrate the inner regions of the cluster more effectively, resulting in more overlap between the BH and NS in the $r \sim 0.05-1~{\rm pc}$ region, where densities are higher. As a result, in general the NSBH GW capture rate is higher in dynammically older GCs than in dynamically younger ones. However, even in the $10$~Gyr GC, the NSBH capture rate is extremely low. Adopting $n_{\rm cl} = 2.9~{\rm Mpc}^{-3}$ for the density of GCs in the universe \citep[e.g.,][]{Portegies+00}, we find volumetric rates of $\Gamma_{\rm NSBH, GC} \sim 10^{-5}~\Gpc^{-3}~\yr^{-1}$ ($7 \times 10^{-5}~\Gpc^{-3}~\yr^{-1}$) for the $1~{\rm Gyr}$ ($10~{\rm Gyr}$) case. This is at least one order of magnitude below every other proposed channels, see Figure \ref{fig:RatesCompilation}. Thus, we conclude that single-single GW capture is not a major contributor to the NSBH merger rate in GCs.

\subsection{Galactic Nuclei (GN)}\label{GN}
It has been shown \citep[e.g.,][]{Bahcall+77} that a spherically symmetric multi-mass stellar population orbiting a SMBH within its radius of influence will relax into a power-law number density profile of the form $n \propto r^{-\alpha}$, where $\alpha$ varies with mass. The most massive members of the cluster will tend to segregate to the center of the cluster. This problem has been studied by various authors using Fokker-Planck formalism with various assumptions about the GN environment \citep[e.g.,][]{Bahcall+77,Freitag+06,Hopman+Alexander06,AH09,Keshet+09,OLeary+09,Aharon+Perets16}. For this work we calculate GW capture rates in GN using BH and NS  density profiles from four different scenarios studied by \citet{Aharon+Perets16} (henceforth AP16). AP16 studied a cluster surrounding a SMBH of mass $4 \times 10^6~\msun$, composed of a two-mass populations of BH ($10~\msun$ \& $30~\msun$), NS ($1.4~\msun$), white dwarfs ($0.6~\msun$), and main-sequence stars ($1~\msun$). We digitally analyzed their Figures 1 and 2 to obtain BH and NS density estimates from the four different GN evolution models they considered: Model 1 --- cluster evolved from pre-existing cusp with compact object (CO) formation in outer cluster regions, does not include $30~\msun$ BH population; Model 2 --- similar to Model 1, but includes the $30~\msun$ BH population; Model 3 --- built-up cluster with in situ star formation in the outer cluster regions; Model 4 --- cluster evolved from pre-existing cusp with CO formation in the inner cluster regions. See AP16 for more details about the assumptions that went into the calculation of these density profiles. We approximate the relative velocity of an encounter at distance $r$ with the circular velocity at distance $r$, as shown in Equation \ref{eq:GNvrel}. We show the cumulative merger rate per GN as a function of $r$ for these four scenarios in Figure \ref{fig:CumuRates}.

We find a total merger rate per GN ranging from $\Gamma_{\rm cl, GN} \sim 6 \times 10^{-11} - 6 \times 10^{-10}~\yr^{-1}$ for our four GN evolution scenarios. The density of GN in the universe is a very uncertain value, but is often assumed in the literature to be in the range of $\sim 0.02-0.04~\Mpc^{-3}$ \citep[e.g.,][]{Conselice+2005,Tsang13}. However, considering a wide range of SMBH masses it may be as high as $\sim 0.1~\Mpc^{-3}$ \citep{Aller+Richstone02,OLeary+09}. Thus, we adopt $n_{\rm cl} = 0.02-0.1~\Mpc^{-3}$ for the cosmic density of GN. We then find a total volumetric rate of NSBH mergers in GN due to sin-sin GW captures of $\Gamma_{\rm NSBH, GN}\sim 0.001-0.06~\volrate$

 As explained in the introduction (also see \citet{OLeary+09} and \citet{Tsang13}), this rate maybe enhanced by a factor of $\xi$, which accounts for the variance in central cusp densities between different GNs. The value of $\xi$ is highly uncertain, but it maybe as high as a few tens \citep{OLeary+09}. Thus, our nominal rate of $\Gamma_{\rm NSBH, GN}\sim 0.001-0.06~\volrate$ is a conservative one.
 
\subsection{Young Stellar Clusters (YSC)}\label{YSC}
 
We perform an order of magnitude estimation adopting models of massive YSCs from \citet{Banerjee17} and \citet{Banerjee18}. We calculate the number density distributions of BHs and NSs, $n_{\rm BH}(r)$ and $n_{\rm NS}(r)$, from the cumulative radial distributions obtained by digitally analyzing the left hand side plots of  Figure 8 from \citet{Banerjee18}. These radial distributions are given for a cluster with initial mass $M_{\rm cl}(t = 0) = 7.5 \times 10^4~\msun$ and metallicity $Z = 0.01~Z_{\odot}$, in snapshots at $t =$ 100 Myr, 1000 Myr, 5000 Myr, 7500 Myr, and 10 Gyr\footnote{The radial distributions are normalized with respect to the total number of bound BH and NS in the cluster, $N_{\rm BH, bound}$ and $N_{\rm NS, bound}$, which are unfortunately given in neither \citet{Banerjee18} nor \citet{Banerjee18} for a $M_{\rm cl}(0) = 7.5 \times 10^4~\msun$ cluster. However, the time evolution of $N_{\rm BH, bound}$ is given for four other cluster masses in the range $M_{\rm cl}(0) = (1-5) \times 10^4~\msun$ in Figure 4 of \citet{Banerjee17}. Based on numbers obtained from \citet{Banerjee17}, figure 4, we fit $N_{\rm BH, bound}(M_{\rm cl})$ with both a linear and quadratic distribution to extrapolate a range for $N_{\rm BH, bound}(M_{\rm cl}(0) = 7.5 \times 10^4)$ at the different time snapshots. $N_{\rm NS, bound}$ is given for a $M_{\rm cl}(0) \approx 3 \times 10^4~\msun$ in Figure 2 of \citet{Banerjee17}, from which we can calculate the ratio $N_{\rm NS, bound}/N_{\rm BH, bound}$ for $M_{\rm cl}(0) \approx 3 \times 10^4~\msun$.  Assuming that this ratio is roughly constant with increasing cluster mass, we can calculate  $N_{\rm NS, bound}(M_{\rm cl}(0) = 7.5 \times 10^4)$ from the extrapolated $N_{\rm BH, bound}$ values. We can now unnormalize $n_{\rm BH}(r)$ and $n_{\rm NS}(r)$ and estimate per cluster sin-sin capture rate.} We assume a cluster velocity dispersion of $3~{\rm km/s}$ for $v_{\rm rel}$ \citep[e.g.,][]{YMCReview}, and single mass distributions of BH and NS of $20~\msun$ and $1.4~\msun$, respectively. We calculate the per cluster merger rate, $\Gamma_{\rm cl,YSC}$, at different time snapshots using Equation \ref{eq:fullrate}. The per cluster merger rate for our nominal YSC model rapidly decreases as the cluster ages, going from $\Gamma_{\rm cl, YSC} \sim 10^{-13}~{\rm yr}^{-1}$ at $t = 100~{\rm Myr}$ to $\Gamma_{\rm cl, YSC} \sim 10^{-16}~{\rm yr}^{-1}$ for $t > 5~{\rm Gyr}$. This is in contrast to what we see with our nominal GC model, where the per cluster rate slightly increases with age. This is primarily due to the fact the half-mass radius of our nominal YSC model increases from $\sim 2$ to $\sim 15$ pc between 100 Myr and 10 Gyr, as shown in Figure 2 of \citet{Banerjee18}. Cluster expansion leads to a decrease in stellar density, which greatly lowers the frequency of dynamical captures. 
 
 We then calculate the volumetric rate similarly to \citet{Ziosi+14}:
 \begin{equation}\label{eq:YSC}
     \Gamma_{\rm NSBH, YSC} \approx \frac{\Gamma_{\rm cl, YSC}}{M_{\rm cl}(0)} ~t_{\rm eff}~\rho_{\rm SF}~f_{\rm SF},
 \end{equation}
where $\rho_{\rm SF} = 1.5 \times 10^{-2}~\msun~\Mpc^{-3}$ is the density of star formation at redshift 0 \citep[adopted from][]{Hopkins+Beacon06}, and $f_{\rm SF} = 0.8$ is the fraction of star formation that takes place in YSCs \citep[e.g.,][]{Lada2x03}. Note that we calculate the locally-detectable rate using the local star-formation density --- as opposed to an integrated cosmological calculation, like that found in \citet{Rastello+20} --- since there is not typically a large time delay between binary formation and binary merger in the sin-sin GW capture channel \citep{OLeary+09,Gondan+18a}. In other words, a NSBH formed via GW capture that merges in the local universe almost certainly formed in the local universe. In light of this, and also since we have found that the per cluster merger rate is strongly dominated by the very early YSC evolution, we consider an ``effective" lifetime for our cluster to be $t_{\rm eff} = 100~{\rm Myr}$ (even though a $7.5 \times 10^4~\msun$ cluster can live up to about 10 Gyr). Thus, we take $\Gamma_{\rm cl, YSC}$ in Equation \ref{eq:YSC} to be $\Gamma_{\rm cl, YSC}(t = 100~{\rm Myr}) \approx 10^{-13}~{\rm yr}^{-1}$. We find $\Gamma_{\rm NSBH, YSC} \approx 2 \times 10^{-3}~\volrate$.
  
We note that this is an upper-limit rate estimation for the following reasons. First of all, our nominal YSC model, with a mass of $7.5 \times 10^4~\msun$\footnote{These relatively more massive young clusters are often referred to in the literature as ``young {\it massive} clusters" (YMCs), see \citet{YMCReview} for a review.}, is much more massive and contain more stellar and compact objects than the average YSC/open cluster (for comparison, the cluster models in \citet{Rastello+20} have masses ranging from $3 \times 10^2-10^3~\msun$). Thus, by using this cluster model as our fiducial model, we are overestimating the per cluster contribution for the average YSC. Secondly, our nominal YSC has a low metallicity of $Z = 0.01~Z_{\odot}$. Since lower cluster metallicity increases the number of compact objects formed, our fiducial cluster metallicity is on the optimistic end of the spectrum. 

We see that our optimistic estimation for the sin-sin merger rate in YSCs is still very low compared to the majority of other merger channels, as seen in Figure \ref{fig:RatesCompilation}. Thus, we can conclude that sin-sin GW capture in YSCs most likely do not contribute to the overall NSBH merger rate.

\section{Discussion}\label{sec:Dis}
\begin{figure}
    \centering
    \includegraphics[width = \linewidth]{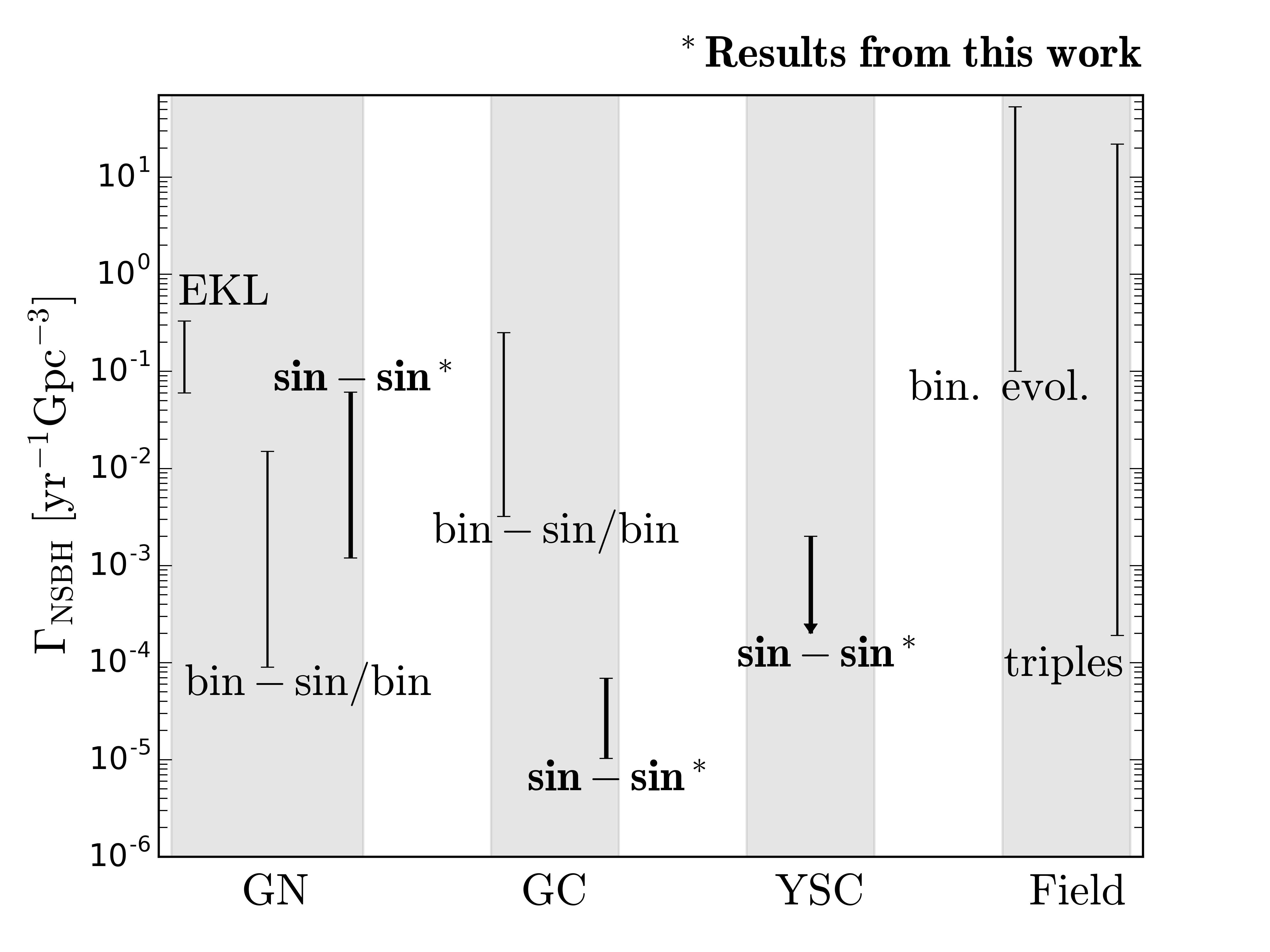}
    \caption{\textbf{Comparison of NSBH merger rates from various channels.} We group merger channels into four major categories: mergers taking place in galactic nuclei (GN), globular clusters (GC), young stellar clusters (YSC), and the galactic field. Within the GN category we highlight three channels: EKL-assisted mergers \citep{Stephan+19}; binary-mediated interactions (bin-sin/bin) \citep{ArcaSedda20}; and sin-sin GW captures (this work). Within the GC category we show rates from binary-mediated interactions (bin-sin/bin) \citep{Clausen+13,Ye+20,ArcaSedda20}; and sin-sin GW captures (this work). Within the YSC category we show the rate from sin-sin GW captures (this work). We denote this rate with an arrow to indicate that this is an upper-limit estimation. Within the Field category we show rates from isolated bin. evol. \citep[iso. bin,][]{Dominik+15,Kruckow+18}; and field triples \citep{Fragione+Loeb19a,Fragione+Loeb19b}.}  
    \label{fig:RatesCompilation}
\end{figure}

In Figure \ref{fig:RatesCompilation}, we compile and compare the predicted NSBH merger rates from the merger channels discussed in Section \ref{sec:Intro}, as well as the results from this work. We group these channels into four major categories:  mergers taking place in GN, GC, YSC, and the galactic field. 
For the GN category we include the following channels: EKL-assisted mergers (rate from \citet{Stephan+19}); binary mediated interactions (rate from \citet{ArcaSedda20}); and sin-sin GW captures (rate compiled from \citet{OLeary+09}, \citet{Tsang13}, and this work). For the GC category we include the following channels: binary mediated interactions (rate compiled from \citet{Clausen+13}, \citet{Ye+20}, and \citet{ArcaSedda20}); and sin-sin GW captures (rate from this work). For the YSC category we include sin-sin GW captures (rate from this work). For the field channel we used the rates from \citet{Dominik+15} and \citet{Kruckow+18}. 

We see from Figure \ref{fig:RatesCompilation} that sin-sin GW captures is highly unlikely to be the dominant mechanism for the production of NSBH mergers. Indeed, sin-sin GW captures do not appear to contribute to the overall NSBH rate in any significant way. The most major caveat to this statement concerns the sin-sin estimate in the GN category. As discussed in Section \ref{sec:Intro}, the sin-sin rates for GN estimated in this work may be underestimated by a ``$\xi$" factor, which is due to the variance of stellar densities in the GN cusp. The actual value of $\xi$ remains highly uncertain --- there are both pessimistic \citep{Tsang13} and optimistic \citep{OLeary+09} estimates of $\xi$ in the literature. If we assume an optimistic value for $\xi$ of a few tens, then the sin-sin GW capture rates in GN will be comparable to both the EKL and field rates. 

In the future, once LIGO has detected a statistical population of NSBH mergers, we will know whether the sin-sin merger rates have been systematically underestimated here by looking at the eccentricity distribution of these mergers. It has been shown that aLIGO can distinguish eccentric stellar-mass compact binary mergers from circular ones for $e \gtrsim 0.05-0.081$ at 10 Hz \citep{Lower+18,Gondan+18c}. Broadly speaking, mergers from dynamical channels are expected to be more eccentric in the LIGO band than mergers from isolated binary evolution in the field, which are expected to be predominantly circular  \citep[e.g.,][]{OLeary+09,Cholis+16,Rodriguez+18,Zevin+18,Lower+18,Samsing18,Gondan+18b,Randall+Xianyu}. Amongst dynamical merger channels, some channels are predicted to yield more eccentric mergers than others. For example, \citet{Rodriguez+18prd} found that roughly 6\% of BBH mergers from bin-sin interactions in GCs have $e \gtrsim 0.05$ in the LIGO band\footnote{\citet{Rodriguez+18prd} distinguishes between bin-sin mergers that take place \textit{after} one or more bin-sin encounters, and mergers that take place \textit{during} a bin-sin encounter due to significant GW emission at close passage. They have termed the latter ``GW capture" mergers. These mergers make up virtually all of their mergers with $e \gtrsim 0.05$ in the LIGO band. While the physical mechanism underlying the GW emission in these ``GW capture" mergers is the same physical mechanism underlying GW emission in sin-sin GW captures, and despite of the similar name, we stress that they are a completely distinct merger channel. For the purposes of this work, we group them with the other bin-sin mergers in GCs. For more information about these mergers, see \citet{Samsing18}, \citet{Rodriguez+18prd}, and \citet{Rodriguez+18}.}. For NSBH mergers from bin-sin encounters in GN, \citet{ArcaSedda20} found that none will have eccentricity above the minimum detection threshold in the LIGO band, but that a large fraction have $e > 0.1$ in the LISA band. Some EKL mergers are also expected to have detectable eccentricities in the LIGO band. Both \citet{Fragione+19b} and \citet{Fragione+Bromberg} found that a non-negligible fraction of BBH EKL mergers will have detectable eccentricity in the LIGO band. However, sin-sin GW capture is by far and away the merger channel that results in the most eccentric mergers in the LIGO band. \citet{Takatsy+Raffai} studied BBH mergers in GN from both EKL and sin-sin GW captures, and found that $\sim 75$\% of sin-sin GW capture mergers will have $e > 0.1$ in the LIGO band, as opposed to only $\sim 10$\% for EKL. Thus, if future NSBH merger observations show a preponderance of eccentric mergers, then it is likely that we have underestimated our sin-sin merger rates here, and most likely in the context of GN. 

The current nominal estimates shown in Figure \ref{fig:RatesCompilation} show four channels that are possible dominant contributors to the NSBH merger rate. These channels have overlapping statistical uncertainties, and are isolated binary evolution, triples in the field, binary-mediated interactions in GCs, and EKL in GN\footnote{Note that the degree of uncertainty in the different channels varies considerably.}. There is a high probability that the future observed NSBH merger population is a ``blend" of two or more merger channels. We may be able to disentangle the contributions from different channels by looking at merger distributions in eccentricity, mass, spin, etc., as different merger channels produce different characteristic distributions in the merger parameter space \citep[e.g.,][]{Rodriguez+18prd,Takatsy+Raffai,ArcaSedda20,Rastello+20}. However, different channels do sometimes overlap in merger parameter space, so it may be very difficult to fully quantify the contribution of each merger channel to the observed distributions. We may also be able to classify individual GW source (although this is probably not be possible for a majority of GW mergers). This can be accomplished with the detection of electromagnetic counterparts \citep[e.g.][]{Lee+10,Tsang13}, or through the detection of imprints on the GW waveform present with some merger channels. For instance, for some EKL-assisted mergers in GN, the gravitational pull of the SMBH on the merging binary can be detected in both LIGO and LISA due to induced GW phase shifts \citep{Inayoshi+17,Meiron+17}, and eccentricity variations \citep{Hoang+19,Randall+Xianyu19,Emami+Loeb20,Deme+20,Gupta+20}.

\section*{Acknowledgements}
We thank Bence Kocsis for helpful discussions. B.M.H. and S.N. acknowledge the partial support of NASA grant No.~80NSSC19K0321 and No.~80NSSC20K0505. S.N. also thanks Howard and Astrid Preston for their generous support. K.K. is supported by an NSF Astronomy and Astrophysics Postdoctoral Fellowship under award AST-2001751.
\bibliography{Binary}
\end{document}